\documentclass[12pt]{article}
\usepackage{amsmath,amssymb,amsfonts,graphicx}

\newcommand{\bnabla}{\mbox{\boldmath $\nabla$}}

\begin{document}

\thispagestyle{empty}
\renewcommand{\refname}{References}

\title{\bf Influence of quantized massive matter fields on the Casimir effect}


\author{Yu. A. Sitenko$^{a,b}$}

\date{}

\maketitle
\begin{center}
$^{a}$ Bogolyubov Institute for Theoretical Physics,\\
National Academy of Sciences of Ukraine,\\
14-b Metrologichna Str., 03680 Kyiv, Ukraine

$^{b}$ Institute for Theoretical Physics, University of Bern,\\
Sidlerstrasse 5, CH-3012 Bern, Switzerland

\end{center}

\begin{abstract}
Charged massive matter fields of spin 0 and 1/2 are quantized in
the presence of an external uniform magnetic field in a spatial
region bounded by two parallel plates. The most general set of
boundary conditions at the plates, that is required by
mathematical consistency and the self-adjointness of the
Hamiltonian operator, is employed. The vacuum fluctuations of the
matter field in the case of the magnetic field orthogonal to the
plates are analyzed, and it is shown that the pressure from the
vacuum onto the plates is positive and independent of the boundary
condition, as well as of the distance between the plates.
Possibilities of the detection of this new-type Casimir effect are
discussed.
\end{abstract}


\bigskip

\begin{center}
\end{center}

\bigskip
\medskip

\section{Introduction}

Almost seven decades ago, Casimir \cite{Cas1} predicted an attraction between grounded metal plates
as a macroscopic effect of vacuum fluctuations in quantum field theory. Since then, his prediction has been confirmed experimentally with great precision, opening prospects for its application in modern nanotechnology, see review in \cite{Bor}.

The detected Casimir force (or pressure) between parallel plates
separated by distance $a$,
$$
 F=-\frac{\pi^{2}}{240}\frac{{\hbar}c}{a^{4}}, \eqno(1)
$$
is due to the vacuum fluctuations of the quantized electromagnetic
field only \cite{Cas1}. As to the vacuum fluctuations of other
quantized fields, their contribution to the Casimir effect was
theoretically considered erstwhile, see, e.g., \cite{Bor}. It
suffices to note here that this contribution is of order of
${\hbar} c a^{-4}$ at $a {\ll} \lambda_{C}$ and of order of
${\hbar}c a^{-4}(a/\lambda_{C})^{\nu}{\exp}(-2a/\lambda_{C})$ at
$a {\gg} \lambda_{C}$, where $\lambda_{C}=\hbar(mc)^{-1}$ is the
Compton wavelength of the matter field of mass $m$; the sign of
this contribution, as well as exponent $\nu$, depends on a
boundary condition and the spin of the matter field. Usually, the
Casimir effect is validated experimentally for the macroscopic
separation of plates: $a > 10^{-8}\,\rm m$.  So, even if one takes
the lightest massive particle, electron ($\lambda_{C} =
3.86\times10^{-13}\,\rm m$), then it becomes clear that the case
of $a {\ll} \lambda_{C}$ has no relation to physics reality.
Whereas, in the realistic case of $a {\gg} \lambda_{C}$, the
contribution of the vacuum fluctuations of quantized massive
matter fields to the Casimir effect is vanishing.

However, quantized massive matter fields can be charged, and as
those perceive an influence from external (classical) electromagnetic fields.
In this letter, I study an impact of an external static uniform
magnetic field on the vacuum of the quantized charged massive
matter fields of spin 0 and 1/2; the matter and external fields
are confined between two parallel plates, with the external field
being orthogonal to the plates. A crucial issue for my analysis is
a choice of boundary conditions at the plates, and I adhere to the
most general one that is required by mathematical consistency and
the self-adjointness of basic operators, see, e.g., \cite{Akhi}.
Namely this circumstance allows us to substantiate a conclusion
that the Casimir effect in the presence of the external magnetic
field orthogonal to the plates is repulsive independently of the
boundary condition and of the distance between the plates.

\section{Choice of boundary conditions}

A relevant basic operator is that of
one-particle energy, which in the case of a spin-0 relativistic
field takes form
$$
H_{(0)}=c\sqrt{-(\hbar{\bnabla})^2+m^2c^2}, \eqno(2)
$$
where ${\bnabla}$ is the covariant derivative involving both
affine and bundle connections. Defining a scalar product as
$(\tilde{\chi},\chi)=\int\limits_{\Omega}{\rm
d}^3r\tilde{\chi}^*\chi$, we get, using integration by parts,
$$
(\tilde{\chi},\,H^2_{(0)}\chi)-(H^2_{(0)}\tilde{\chi},\chi)=-{\rm
i}\int\limits_{\partial\Omega}{\rm d}\boldsymbol{\sigma}\cdot{\bf
J}[\tilde{\chi},\chi], \eqno(3)
$$
where $\partial\Omega$ is a two-dimensional surface bounding the three-dimensional spatial region $\Omega$, and
$$
{\bf J}[\tilde{\chi},\chi]=-{\rm i}(\hbar
c)^2[\tilde{\chi}^*({\bnabla}\chi)-({\bnabla}\tilde{\chi})^*\chi].\eqno(4)
$$
The squared Hamiltonian operator, $H^2_{(0)}$, is Hermitian (or
symmetric in mathematical parlance) if the right-hand side of (3)
vanishes, or
$$
{\bf n}\cdot{\bf J}[\tilde{\chi},\chi]|_{{\bf r}\in
\partial\Omega}=0,\eqno(5)
$$
where $\boldsymbol{n}$ is the unit normal to the
boundary. The latter condition can be satisfied in various ways by
imposing different boundary conditions for $\chi$ and
$\tilde{\chi}$. However, among the whole variety, there may exist
a possibility that a boundary condition for $\tilde{\chi}$ is the
same as that for $\chi$; then operator $H^2_{(0)}$ is
self-adjoint.

In the context of the Casimir effect, one usually considers a disconnected noncompact boundary consisting of two
connected components, $\partial\Omega=\partial\Omega^{(+)}\bigcup\partial\Omega^{(-)}$. Introducing coordinates ${\bf r}=(x,y,z)$
in such a way that $x$ and $y$ are tangential to the boundary, while $z$ is normal to it, one identifies the position
of $\partial\Omega^{(\pm)}$ with, say, $z=\pm a/2$. Then condition (5) takes form
$$
J^z[\tilde{\chi},\chi]|_{z=a/2}-J^z[\tilde{\chi},\chi]|_{z=-a/2}=0.\eqno(6)
$$
The left-hand side of (6) can be presented in the form
\newpage
$$
J^z[\tilde{\chi},\chi]|_{z=a/2}-J^z[\tilde{\chi},\chi]|_{z=-a/2}=\frac{(\hbar
c)^2}{2a}
$$
$$
\times\left\{[(\tilde{\chi}+{\rm
i}a\nabla_z\tilde{\chi})^*(\chi+{\rm
i}a\nabla_z\chi)]|_{z=-a/2}+[(\tilde{\chi}-{\rm
i}a\nabla_z\tilde{\chi})^*(\chi-{\rm
i}a\nabla_z\chi)]|_{z=a/2}\right.
$$
$$
\left.-[(\tilde{\chi}-{\rm i}a\nabla_z\tilde{\chi})^*(\chi-{\rm
i}a\nabla_z\chi)]|_{z=-a/2}-[(\tilde{\chi}+{\rm
i}a\nabla_z\tilde{\chi})^*(\chi+{\rm
i}a\nabla_z\chi)]|_{z=a/2}\right\}. \eqno(7)
$$
It is obvious that (6) is satisfied if the following condition holds:
$$
\begin{pmatrix} (\chi+{\rm i}a\nabla_z\chi)|_{z=-a/2} \\ (\chi-{\rm i}a\nabla_z\chi)|_{z=a/2}\end{pmatrix}=
U\begin{pmatrix} (\chi-{\rm i}a\nabla_z\chi)|_{z=-a/2} \\
(\chi+{\rm i}a\nabla_z\chi)|_{z=a/2}\end{pmatrix} \quad \eqno(8)
$$
(with the same condition holding for $\tilde{\chi}$), where $U$ is
a $U(2)$-matrix which is in general parametrized as
$$
U={\rm e}^{-{\rm i}\mu}\begin{pmatrix}
u&v\\
-v^{*}&u^{*}
\end{pmatrix},\quad 0\leq\mu<\pi,\quad |u|^2+|v|^2=1. \eqno(9)
$$
Thus, the squared Hamiltonian operator, $H^2_{(0)}$, is
self-adjoint under condition (8), and four real continuous
parameters from (9) can be interpreted as the self-adjoint
extension parameters.

However, our quest is for $H_{(0)}$ itself to be self-adjoint, and this demand diminishes the number of the
self-adjoint extension parameters. Let $\psi({\bf r})$ be a solution to the stationary Klein-Fock-Gordon equation
$$
H^2_{(0)}\psi=\hbar^2\omega^2\psi.\eqno(10)
$$
If the spectrum is continuous, $\hbar^2\omega^2\geq m^2c^4$,
meaning that $(\psi,H^2_{(0)}\psi)\geq m^2c^4(\psi,\psi)$,
then $U^{\dagger}=U$, resulting in constraint: $\mu=\pi/2$, $u^*=-u$,
which reduces the number of the self-adjoint extension parameters by half. A further physical restriction is by imposing condition
$$
J^z[\tilde{\chi},\chi]|_{z=\pm a/2}=0\eqno(11)
$$
rather than condition (6); this means that the matter is confined
within the boundaries. Then $U^*U=I$, and one more constraint is
added, $v^*=-v$,
resulting in the real $U(2)$-matrix,
$$
U=\sigma^1\cos\rho+\sigma^3\sin\rho, \quad 0\leq\rho<2\pi,\eqno(12)
$$
with parametrization ${\rm Im}u=\sin\rho$, ${\rm Im}v=\cos\rho$,
employing only one self-adjoint extension parameter, $\rho$,
\cite{Si1}. The above restrictions ensure the self-adjointness of
the Hamiltonian operator, $H_{(0)}$, under the boundary condition
taking explicitly the form
$$
\left\{\begin{array}{l}
    \chi|_{z=-a/2}=\tan\left(\frac 12\rho+\frac \pi4\right)\chi|_{z=a/2} \\ [3 mm]
    \nabla_z\chi|_{z=-a/2}={\rm cot}\left(\frac 12\rho+\frac \pi4\right)\nabla_z\chi|_{z=a/2}
\end{array}\right\}
\eqno(13)
$$
(the same condition is for $\tilde{\chi}$).

Clearly, operator $H_{(0)}$ is self-adjoint under the Dirichlet
($U=-I$) or Neumann ($U=I$) boundary condition as well. The mixed,
Dirichlet-Neumann ($U=-\sigma^3$) or Neumann-Dirichlet
($U=\sigma^3$), boundary condition is obtainable as a particular
case of (13) at $\rho=3\pi/2$ or $\rho=\pi/2$, respectively; the
periodic ($U=\sigma^1$) and
 antiperiodic ($U=-\sigma^1$) boundary conditions correspond to $\rho=0$ and $\rho=\pi$.

Turning now to the case of a spin-1/2 relativistic field, a
relevant basic operator is that of the Dirac Hamiltonian:
$$
H_{(1/2)}=c(-{\rm i}\hbar\boldsymbol{\alpha}\cdot{\bnabla}+\beta
mc). \eqno(14)
$$
Defining a scalar product as
$(\tilde{\chi},\chi)=\int\limits_{\Omega}{\rm
d}^3r\tilde{\chi}^{\dagger}\chi$, we get, using integration by
parts,
$$
(\tilde{\chi},H_{(1/2)}\chi)-(H_{(1/2)}\tilde{\chi},\chi)=-{\rm
i}\int\limits_{\partial\Omega}{\rm d}\boldsymbol{\sigma}\cdot{\bf
J}[\tilde{\chi},\chi], \eqno(15)
$$
where
$$
{\bf J}[\tilde{\chi},\chi]=\hbar
c\tilde{\chi}^{\dagger}\boldsymbol{\alpha}\chi \eqno(16)
$$
in this case. The Dirac operator is Hermitian if condition (5)
holds, while it is self-adjoint when this condition is resolved by
the same boundary conditions for $\chi$ and $\tilde{\chi}$.
Considering a disconnected noncompact boundary consisting of two
connected components, let us restrict ourselves to the case of the
matter which is confined within the boundaries, see (11). Then the
most general condition ensuring the self-adjointness of operator
$H_{(1/2)}$ involves four self-adjoint extension parameters
\cite{Si2}:
$$
[\beta -
\frac{I(\cosh^2\tilde{\vartheta}_{\pm}+1)-\beta\sinh^2\tilde{\vartheta}_{\pm}}{2{\rm
i}\cosh\tilde{\vartheta}_{\pm}}({\pm}{\alpha}^z\cosh\vartheta_{\pm}+\beta\gamma^{5}\sinh\vartheta_{\pm})]\chi|_{z=\pm{a/2}}=0
\eqno(17)
$$
(the same condition is for $\tilde{\chi}$), where ${\alpha}^z$ is
the component of $\boldsymbol{\alpha}$ in the direction of the
$z$-axis, and $\gamma^5={\rm i}\alpha^1\alpha^2\alpha^3$.

It should be noted that, in general, the values of the
self-adjoint extension parameters both in the spin-0 and spin-1/2
cases may vary arbitrarily from point to point of the boundary
surface. However, such a generality seems to be excessive,
moreover, it is impermissible, as long as boundary conditions (8),
(13) and (17) are to be regarded as the ones determining the
spectrum of the wave number vector in the $z$-direction.
Therefore, it is assumed in the following that the self-adjoint
extension parameters are independent of coordinates $x$ and $y$.

In the spin-1/2 case, any immediate physical motivation to diminish the number
of the self-adjoint extension parameters seems to be lacking. In this situation
one can be guided by such arguments as simplicity and unambiguity of the determination
of the spectrum of $k_l$ -- $z$-component of the wave number vector. In particular, the
condition that this spectrum be independent of the values of other components of the wave
number vector yields restriction
$$
\vartheta_{+}=\vartheta_{-}=\vartheta, \quad
\tilde{\vartheta}_{+}=\tilde{\vartheta}_{-}=0, \eqno(18)
$$
with resulting boundary condition
$$
(\beta \pm {\rm i}\alpha^{z}\cosh\vartheta +{\rm
i}\beta\gamma^{5}\sinh\vartheta)\chi|_{z=\pm{a/2}}=0 \eqno(19)
$$
(and the same condition for $\tilde{\chi}$). The standard MIT bag
boundary condition (see, e.g., \cite {Joh}) corresponds to
$\vartheta=0$.

In the spin-0 case, the spectrum of $k_l$ is determined by condition
$$
\sin\left[\frac{1}{2}(k_{l}a+\rho)\right]=0 \quad (-\infty<k_{l}<\infty), \quad \rho\neq{\pi}/{2},\,3{\pi}/{2},\eqno(20)
$$
or
$$
\cos(k_{l}a)=0 \quad (0<k_{l}<\infty), \quad \rho={\pi}/{2},\,3{\pi}/{2},\eqno(21)
$$
stemming from (13). In the spin-1/2 case, the spectrum of $k_l$ is determined by condition
$$
\sin\left[k_{l}a+{\rm arctan}\left(\frac{\hbar
k_{l}\cosh\vartheta}{mc}\right)\right]=0 \quad (0<k_{l}<\infty),
\eqno(22)
$$
stemming from (19).

    \section{Casimir force}

 The operator of a charged massive spin-0 field which is quantized in a static background is presented in the
form
$$\hat{\Psi}_{(0)}(t,\mathbf{r})=\sum\!\!\!\!\!\!\!\!\!\int \sqrt{\frac{c}{2\omega}}\left[{\rm e}^{-{\rm i}\omega t}\psi
(\mathbf{r}) \hat{a} + {\rm e}^{{\rm i}\omega
t}\psi^*(\mathbf{r})\hat{b}^{\dag}\right],\eqno(23)
$$
where $\hat{a}^{\dag}$ and $\hat{a}$ ($\hat{b}^{\dag}$ and
$\hat{b}$) are the spin-0 particle (antiparticle) creation and
destruction operators satisfying commutation relations, symbol
$\sum\!\!\!\!\!\!\int$ denotes summation over discrete and
integration (with a certain measure) over continuous values of the
components of the wave number vector, and wave functions
$\psi(\textbf{r})$ form a complete set of solutions to the
stationary Klein-Fock-Gordon equation, see (10).

There is some arbitrariness in the definition of the energy-momentum tensor for bosonic fields in flat space-time: one can add
term $\xi\nabla_{\rho}\Xi^{\mu\nu\rho}$, where $\Xi^{\mu\nu\rho}=-\Xi^{\mu\rho\nu}$, to the canonically-defined energy-momentum
tensor, $T^{\mu\nu}_{\rm can}$, see, e.g., \cite{Itz}. For instance, the whole construction for a spin-0 field at $\xi=1/6$ is known
as the improved energy-momentum tensor which is adequate for the implementation of conformal invariance in the $m=0$ case \cite{Che,Cal}.
Hence, the temporal component of the operator of the energy-momentum tensor for the spin-0 field is in general
$$
\hat{T}^{00}_{(0)}=\frac{\hbar}{c}\left\{[{\partial_t}\hat{\Psi}^{\dag},{\partial_t}\hat{\Psi}]_{+}-
\left[\frac{1}{4}{\partial_t}^2-c^2\left(\frac{1}{4}-\xi\right)\bnabla^2\right]
[\hat{\Psi}^{\dag},\hat{\Psi}]_{+}\right\},\eqno(24)
$$
and the vacuum expectation value of the energy density is given by the following formal expression
$$
\varepsilon_{(0)}\equiv
<\rm{vac}|\hat{T}^{00}_{(0)}|\rm{vac}>=\hbar
\sum\!\!\!\!\!\!\!\!\!\int\omega\psi^{*}(\textbf{r})\psi(\textbf{r})
$$
$$
+\left(\frac{1}{4}-\xi\right)\hbar
c^2\sum\!\!\!\!\!\!\!\!\!\int\omega^{-1}\bnabla^2\psi^{*}(\textbf{r})\psi(\textbf{r});\eqno(25)
$$
however, physical observables should be certainly independent of the value of $\xi$.

The operator of a spin-1/2 field which is quantized in a static
background is presented in the form
$$\hat{\Psi}_{(1/2)}(t,\mathbf{r})=\sum\!\!\!\!\!\!\!\!\!\!\int\limits_{\omega>0}{\rm e}^{-{\rm i}\omega t}\psi(\mathbf{r})\hat{a}
+\sum\!\!\!\!\!\!\!\!\!\!\int\limits_{\omega<0}{\rm e}^{-{\rm
i}\omega t}\psi(\mathbf{r})\hat{b}^{\dag},\eqno(26)
$$where
$\hat{a}^{\dag}$ and $\hat{a}$ ($\hat{b}^{\dag}$ and $\hat{b}$)
are the spin-1/2 particle (antiparticle) creation and destruction
operators satisfying anticommutation relations, and wave functions
$\psi(\textbf{r})$ form a complete set of solutions to the
stationary Dirac equation
$$
H_{(1/2)}\psi(\mathbf{r})=\hbar \omega \psi(\mathbf{r}).\eqno(27)
$$
The temporal component of the operator of the energy-momentum
tensor is given by expression
$$
\hat{T}^{00}_{(1/2)}=\frac{\rm{i}\hbar}{4}[\hat{\Psi}^{\dag}({\partial_t}\hat{\Psi})-({\partial_t}\hat{\Psi}^{T})
\hat{\Psi}^{{\dag}T}-({\partial_t}\hat{\Psi}^{\dag})\hat{\Psi}+\hat{\Psi}^{T}({\partial_t}\hat{\Psi}^{{\dag}T})],\eqno(28)
$$where superscript $T$ denotes a transposed spinor. Consequently, the formal expression for the vacuum expectation value of the energy
density is
$$
\varepsilon_{(1/2)}\equiv<\rm{vac}|\hat{T}^{00}_{(1/2)}|\rm{vac}>=-\frac{\hbar}{2}\sum\!\!\!\!\!\!\!\!\!\int
\,|\omega|\psi^{\dag}(\textbf{r})\psi(\textbf{r}).\eqno(29)
$$

Let us consider the quantization of the charged massive field in
the background of a uniform magnetic field $(\mathbf{B})$ in flat
space-time, then the covariant derivative is defined as
$$\bnabla\hat{\Psi}=\left(\boldsymbol{\partial}-\frac{{\rm i}e}{\hbar
c}\mathbf{A}\right)\hat{\Psi},\quad
\bnabla\hat{\Psi}^{\dag}=\left(\boldsymbol{\partial}+\frac{{\rm
i}e}{\hbar c}\mathbf{A}\right)\hat{\Psi}^{\dag} \quad
(\mathbf{B}=\boldsymbol{\partial}\times\mathbf{A}),\eqno(30)
$$
$e$ is the particle charge and the gauge can be chosen as
$\mathbf{A}=(-yB,0,0)$, where $B$ is the magnetic field strength
which is directed along the $z$-axis in Cartesian coordinates
$\mathbf{r}=(x,y,z)$. The formal expression for the vacuum energy
density in this case is readily obtained:
$$
\varepsilon^{\infty}_{(s)}=\frac{|eB|}{(2\pi)^{2}c}(1-4s)\int\limits_{-\infty}^{\infty}{\rm
d}k\sum\limits_{n=0}^{\infty}
(1+2s-2s\delta_{n0})|\omega_{snk}|,\quad s=0,1/2,\eqno(31)
$$
where the superscript on the left-hand side indicates that the external magnetic field fills the whole (infinite) space, and the one-particle energy spectrum is
$$
\hbar|\omega_{snk}|=\sqrt{|eB|\hbar c(2n+1-2s)+\hbar^{2}
c^{2} k^{2}+m^{2}c^{4}},
$$
$$
-\infty<k<\infty, \quad n=0,1,2,...,\,\,\, \eqno(32)
$$
$k$ is the value of the wave number vector along the $z$-axis, $n$
labels the Landau levels. It should be noted that dependence on
$\xi$ in the $s=0$ case has disappeared (just due to vanishing
$\sum\!\!\!\!\!\!\int\omega^{-1}\bnabla^2\psi^{*}(\textbf{r})\psi(\textbf{r})$).
The integral and the sum in (31) are divergent and require
regularization and renormalization. This problem has been solved
long ago by Weisskopf \cite{Wei} (in the $s=0$ case) and
Heisenberg and Euler \cite{Hei2} (in the $s=1/2$ case), see also
\cite{Schw} and review in \cite{Dun}, and we just list here their
result:
$$\varepsilon^{\infty}_{(s)\rm ren}=\frac{e^{2}B^{2}}{{(4\pi)}^{2}\hbar c}\int\limits_{0}^{\infty}\frac{{\rm d}\eta}{\eta}\,{\rm exp}\left(-\frac{m^{2}c^{3}\eta}{\hbar |eB|}\right)
\biggl[\frac{4s\cosh\eta - 1 + 2s}{\eta\sinh\eta}
$$
$$
 + (1-6s)\frac{1}{{\eta}^{2}} - \frac{1}{6}(1+6s)\biggr];\eqno(33)
$$
note that the renormalization procedure involves subtraction at
$B=0$ and renormalization of the charge.

Let us turn now to the quantization of the charged massive field in
the background of a static uniform magnetic field in spatial
region $\Omega$ bounded by two parallel surfaces $\partial{\Omega}^{(+)}$
and $\partial{\Omega}^{(-)}$; the position of $\partial{\Omega}^{(\pm)}$ is identified
with $z=\pm{a/2}$, and the magnetic field is orthogonal to the
boundary. In the $s=0$ case, we take account for relation
$$
\int\limits_{-a/2}^{a/2}{\rm{d}}z\,\bnabla^2\psi^{*}(\textbf{r})\psi(\textbf{r})=[\left(\nabla_z\psi^{*}\right)\psi]|_{z=-{a/2}}^{z={a/2}}
+ [\psi^{*}\left(\nabla_z\psi\right)]|_{z=-{a/2}}^{z={a/2}},\eqno(34)
$$
and note that its right-hand side vanishes under the one-parameter boundary condition given by (13), as well as the Dirichlet and Neumann boundary conditions
(but not under the more general four-parameter boundary condition given by (8)). Thus, by imposing conditions (13) and (19) in the $s=0$ and $s=1/2$ cases, respectively,
we obtain the following formal expression for the vacuum expectation value of the energy per unit area of the boundary surface
$$
\frac{E_{(s)}}{L^2}\equiv\int\limits_{-a/2}^{a/2}{\rm{d}}z\,\varepsilon_{(s)}=\frac{|eB|}{2\pi c}(1-4s)\sum\limits_{l}\sum\limits_{n=0}^{\infty}
(1+2s-2s\delta_{n0})|\omega_{snk_l}|,\eqno(35)
$$
which is $\xi$-independent in the $s=0$ case; the discrete spectrum of $k_l$ is determined by conditions (20) or (21) and (22) in the $s=0$ and $s=1/2$ cases, respectively, and
$L$ is a length characterizing the area of the boundary surface ($L\rightarrow\infty$).

As was already mentioned, the expression for the induced vacuum
energy per unit area of the boundary surface, see (35), can be
regarded as purely formal, since it is ill-defined due to the
divergence of infinite sums over $l$ and $n$. To tame the
divergence, a factor containing a regularization parameter should
be inserted in (35). A  summation over values of $k_{l}$, which
are determined by (20)-(22), is to be performed  with the use of
versions of the Abel-Plana formula, that were derived in \cite
{Si1,Si2,Bel1,Bel2}. According to these versions, a contribution
of the boundaries is separated into a piece which is free from
divergences, and, therefore, the regularization can be safely
removed in this piece. The remaining (divergent) piece consists of
two terms: one is equal to $\varepsilon^{\infty}_{(s)}$ (31)
multiplied by $a$, and another one is independent of $a$.
Employing the same renormalization procedure as in the case of no
boundaries, when the magnetic field fills the whole space, we
determine the Casimir energy, ${E_{(s)\rm ren}}/{L^2}$, by
substituting $\varepsilon^{\infty}_{(s)\rm ren}$ (33) for
$\varepsilon^{\infty}_{(s)}$. As to the $a$-independent divergent
term (which can be interpreted as describing the proper energies
of the boundary surfaces), it is of no concern for us, since, rather than the Casimir
energy, a physically relevant quantity is the Casimir force which
is defined as
$$
F_{(s)}=-\frac{\partial}{\partial a}\frac{E_{(s)\rm ren}}{L^2},\eqno(36)
$$
and which is thus free from divergences. In this way, we obtain
$$
F_{(s)}=-\varepsilon^{\infty}_{(s)\rm ren}
$$
$$
-\frac{|eB|}{\pi^{2}}\sum\limits_{n=0}^{\infty}(1+2s-2s\delta_{n0})\int\limits_{M_{sn}c/{\hbar}}^{\infty}{\rm
d}\kappa
\Upsilon_{(s)}(\kappa){\kappa}^{2-4s}(\kappa^{2}-M_{sn}^{2}c^{2}/{\hbar}^{2})^{2s-1/2},\eqno(37)
$$
where
$$
M_{sn}=\sqrt{|eB|\hbar c^{-3}(2n+1-2s)+m^{2}},\eqno (38)
$$
$$
\Upsilon_{(0)}(\kappa)=\frac{1}{2}\,\frac{\cos \rho-{\rm
e}^{-{\kappa}a}}{{\rm cosh}({\kappa}a)-\cos\rho} \eqno(39)
$$
and
\newpage
$$
\Upsilon_{(1/2)}(\kappa) =\biggl\{\left[\left(2\kappa a-1\right)
\left(\kappa^{2}\cosh^{2}\vartheta-m^{2}c^{2}{\hbar}^{-2}\right)-2{\kappa}mc{\hbar}^{-1}\cosh\vartheta\right]
{\rm e}^{2{\kappa}a}
\biggr.
$$
$$
\biggl.-\left(\kappa\cosh\vartheta-mc{\hbar}^{-1}\right)^{2}\biggr\}
\biggl[\left(\kappa\cosh\vartheta+mc{\hbar}^{-1}\right){\rm
e}^{2{\kappa}a}+\kappa\cosh\vartheta-mc{\hbar}^{-1}\biggr]^{-2}.\eqno(40)
$$

For a particular choice of the boundary condition yielding spectrum
 $k_{l}=\frac{\pi}{a}(l+\frac{1}{2})$ $(l=0,1,2,...)$,
we obtain
$$
\left\{\begin{array}{l}\left.F_{(0)}\right|_{\rho={\pi}/{2},\,3{\pi}/{2}}
\\
\left.F_{(1/2)}\right|_{\vartheta=\pm\infty} \end{array} \right\}
= -\varepsilon^{\infty}_{(s)\rm ren}
$$
$$
+\frac{|eB|}{\pi^{2}}(1-4s)\sum\limits_{n=0}^{\infty}(1+2s-2s\delta_{n0})\int\limits_{M_{sn}c/\hbar}^{\infty}\frac{{\rm
d}\kappa}{{\rm
e}^{2{\kappa}a}+1}\frac{\kappa^{2}}{\sqrt{\kappa^{2}-M_{sn}^{2}c^{2}/\hbar^{2}}}.\eqno(41)
$$
It should be recalled that $\rho={\pi}/{2}$ or $\rho=3{\pi}/{2}$ in the $s=0$ case corresponds to the mixed (Neumann-Dirichlet or Dirichlet-Neumann) boundary condition.
The Casimir force in the $s=0$ case under either Dirichlet or Neumann boundary condition, when the spectrum is $k_{l}=\frac{\pi}{a}l$ $(l=1,2,...)$, is given by expression
$$
F_{(0)}=-\varepsilon^{\infty}_{(0)\rm ren}-
\frac{|eB|}{\pi^{2}}\sum\limits_{n=0}^{\infty}\int\limits_{M_{0n}c/\hbar}^{\infty}\frac{{\rm
d}\kappa}{{\rm
e}^{2{\kappa}a}-1}\frac{\kappa^{2}}{\sqrt{\kappa^{2}-M_{0n}^{2}c^{2}/\hbar^{2}}}.\eqno(42)
$$
By changing $a\rightarrow{a/2}$ in (42) we obtain the Casimir force in the $s=0$ case under the periodic boundary condition,
when the spectrum is $k_{l}=\frac{2\pi}{a}l$ $(l=0,\pm 1,\pm 2,...)$. By changing $a\rightarrow{a/2}$ in (41) at $s=0$ we obtain
the Casimir force in the $s=0$ case under the antiperiodic boundary condition, when the spectrum is $k_{l}=\frac{2\pi}{a}(l+\frac{1}{2})$
$(l=0,\pm 1,\pm 2,...)$.

Note also that the antiperiodic boundary condition in the $s=1/2$ case,
$$
\chi|_{z=-a/2}+\chi|_{z=a/2}=0 \eqno(43)
$$
(the same condition is for $\tilde{\chi}$ ), ensures the
self-adjointness of the Dirac Hamiltonian operator, but the matter
is not confined within the boundaries: instead, the influx of the
matter at one boundary surface equals the outflux of the matter at
another boundary surface (condition (6) holds instead of condition
(11)). By changing $a\rightarrow{a/2}$ in (41) at $s=1/2$ we
obtain the Casimir force in the case of (43), when the spectrum is
$k_{l}=\frac{2\pi}{a}(l+\frac{1}{2})$ $(l=0,\pm1,\pm2,...)$.

\section{Conclusion}

The influence of an external uniform magnetic field and boundary
conditions on the vacuum of quantized charged matter fields (of
mass $m$ and spin $s=0, \, 1/2$) confined between two parallel
plates has been comprehensively analyzed, and the Casimir force
acting onto the plates is found to take the form, see (37):
$$
F_{(s)}=-\varepsilon^{\infty}_{(s)\rm ren}-f_{(s)}(a), \eqno(44)
$$
where all dependence on the distance ($a$) between the plates
and the choice of a boundary condition is contained in the second
term, $-f_{(s)}(a)$. In the physically meaningful case, $amc/{\hbar} \gg 1$,
this second term is exponentially damped as $\exp(-2amc/{\hbar})$,
and the Casimir force is given by the first term,
$F_{(s)}=-\varepsilon^{\infty}_{(s)\rm ren}$. It should be noted
that the Heisenberg-Euler-Weisskopf vacuum energy density,
$\varepsilon^{\infty}_{(s)\rm ren}$, see (33), is negative (vanishing at
$B=0$ only), hence, the Casimir effect is repulsive, i.e. the pressure
from the vacuum onto the plates is positive.  Defining the critical value
of the magnetic field as $B_{\rm crit}={m^2 c^3(\hbar}|e|)^{-1}$, one
can obtain in the limit of a supercritical magnetic field,
$|B|{\gg}B_{\rm crit}$, from (33):
$$
F_{(s)}=\frac{1}{24\pi^{2}}\left[1-\frac{3}{2}\left(\frac{1}{2}-s\right)\right]\frac{{\hbar}c}{\lambda_{C}^{4}}\left(\frac{B}{B_{\rm
crit}}\right)^{2}\ln\frac{2|B|}{B_{\rm crit}}\eqno (45)
$$
(recall that $\lambda_{C}=\hbar(mc)^{-1}$ is the Compton wavelength).
Note that the critical value is the lowest one, $B_{\rm crit}=4.41\times 10^{13}\,\rm G$, 
for the case of quantized electron-positron matter, and supercritical
magnetic fields with $|B|  \gg 10^{13}\,\rm G$ may
be attainable in some astrophysical objects, such as neutron stars
and magnetars \cite{Har}, and also gamma-ray bursters in scenarios
involving protomagnetars \cite{Met}. A proper account for the
influence of Casimir pressure (45) on physical processes in
these objects should be taken.

Supercritical magnetic fields are not feasible in terrestrial laboratories
where the maximal values of steady magnetic fields are of order of $10^{5}\,\rm G$,
see, e.g., \cite{Per}. In the case of a subcritical magnetic field,
$|B|{\ll}B_{\rm crit}$, one obtains from (33):
$$
F_{(s)}=\frac{1}{360\pi^{2}}\left[1-\frac{9}{8}\left(\frac{1}{2}-s\right)\right]\frac{{\hbar}c}{\lambda_{C}^{4}}\left(\frac{B}{B_{\rm
crit}}\right)^{4}.\eqno (46)
$$
Let us compare this with the attractive Casimir force which is due
to the quantized electromagnetic field, see $F$ (1), and
define ratio
$$
\frac{F_{(s)}}{F}=-\frac{2}{3\pi^{4}}\left[1-\frac{9}{8}\left(\frac{1}{2}-s\right)\right]\left(\frac{a}{\lambda_{C}}\right)^{4}\left(\frac{B}{B_{\rm
crit}}\right)^{4}. \eqno (47)
$$
At $a=10^{-6}\,\rm m$ and $B=10^{5}\,\rm G$ the attraction is prevailing
over the repulsion by six orders of magnitude,
$F/F_{(s)}\approx-10^{6}$, and the Casimir force is
$F\approx-1.3\,\rm mPa$. However, at $a=10^{-5}\,\rm m$ and
$B=10^{6}\,\rm G$ the repulsion becomes dominant over the attraction
by two orders of magnitude, $F_{(s)}/F\approx-10^{2}$ and the Casimir
force in the $s=1/2$ case takes value $F_{(1/2)} \approx 0.009\,\rm mPa$. Otherwise, the same
value of the Casimir force is achieved at $a=10^{-6}\,\rm m$ and
$B=10^{7}\,\rm G$. Thus, an experimental observation of the influence
of the external magnetic field on the Casimir pressure seems to be
possible in some future in terrestrial laboratories.

\section*{Acknowledgments}

I would like to thank P.~Minkowsky, F.~Niedermayer and U.-J.~Wiese
for fruitful discussions and interesting remarks. The research was
supported by the National Academy of Sciences of Ukraine (project
No. 0112U000054). A partial support from the Program of Fundamental
Research of the Department of Physics and Astronomy of the
National Academy of Sciences of Ukraine (project No. 0112U000056)
and from the ICTP -- SEENET-MTP grant PRJ-09 ``Strings and
Cosmology'' is also acknowledged.


\begin{thebibliography}{0}

\bibitem{Cas1}  H.B.G.Casimir, {\it Proc. Kon. Ned. Akad. Wetenschap
B} {\bf 51}, 793 (1948),

\bibitem{Bor} M.Bordag, G.L.Klimchitskaya, U.Mohideen and V.M.Mostepanenko,
{\it Advances in the Casimir Effect} (Oxford University Press, Oxford, 2009).

\bibitem{Akhi}
N.I.Akhiezer and I.M.Glazman, {\it Theory of Linear Operators in Hilbert
Space} (Pitman, Boston, 1981).

\bibitem{Si1}
Yu.A.Sitenko and S.A.Yushchenko, {\it Intern. J. Mod. Phys. A} {\bf 29}, 1450052 (2014).

\bibitem{Si2}
Yu.A.Sitenko, {\it Phys. Rev. D} {\bf 91}, 085012 (2015).

\bibitem{Joh}
K.Johnson, {\it Acta  Phys. Pol. B} {\bf 6}, 865 (1975).

\bibitem{Itz} C.Itzykson and J.-B.Zuber, {\it Quantum Field Theory} (Dover, New York, 2005).

\bibitem{Che} N.A.Chernikov and E.A.Tagirov, {\it Ann. Inst. Henri
Poincare A} \textbf{9}, 109 (1968).

\bibitem{Cal} C.G.Callan, S.Coleman and R.Jackiw, {\it Ann. Phys. (N.Y.)}
\textbf{59}, 42 (1970).

\bibitem{Wei}
V.S.Weisskopf, {\it Kong. Dans. Vid. Selsk. Mat-Fys. Medd.} {\bf 14}, 6 (1936).

\bibitem{Hei2}
W.Heisenberg and H.Euler, {\it Z. Phys.} {\bf 98}, 714 (1936).

\bibitem{Schw}
J.Schwinger, {\it Phys. Rev.} {\bf 82}, 662 (1951).

\bibitem{Dun}
G.V.Dunne, 'Heisenberg-Euler effective lagrangians: Basics and
extensions'. In: {\it Ian Kogan Memorial Collection 'From Fields
to Strings: Circumnavigating Theoretical Physics'.} Ed. by
M.Shifman, A.Vainshtein and J.Wheater (World Scientific,
Singapore, 2004) Vol.1, pp. 445-522.

\bibitem{Bel1}
S.Bellucci and A.A.Saharian, {\it Phys. Rev. D} {\bf 80}, 105003
(2009).

\bibitem{Bel2}
S.Bellucci, A.A.Saharian and V.M.Bardeghyan, {\it Phys. Rev. D} {\bf 82}, 065011 (2010).

\bibitem{Har}
A.K.Harding and D.Lai, {\it Rep. Progr. Phys.} {\bf 69}, 2631 (2006).

\bibitem{Met}
B.D.Metzger, D.Giannios, T.A.Thompson, N.Bucciantini and
E.Quataert, {\it Mon. Not. Roy. Astron. Soc.} {\bf 413}, 2031
(2011).

\bibitem{Per}%
J.A.A.J.Perenboom, J.C.Maan, M.R.van Breukelen, S.A.J.Wiegers, A.den Ouden, C.A.Wulfers, W.J.van der Zande, R.T.Jongma, A.F.G.van der Meer and B.Redlich,
{\it J. Low Temp. Phys.} \textbf{170}, 520 (2013).


\end{thebibliography}
\end{document}